\documentclass[prl,showpacs,showkeys,nofootinbib,twocolumn]{revtex4}

\usepackage{graphicx}  %
\usepackage{bm}  %

\newcommand{\beq}{\begin{equation}}
\newcommand{\eeq}{\end{equation}}
\newcommand{\bqa}{\begin{eqnarray}}
\newcommand{\eqa}{\end{eqnarray}}

\def\square{\vcenter{\vbox{\hrule height.4pt
          \hbox{\vrule width.4pt height8pt
          \kern8pt\vrule width.4pt}\hrule height.4pt}}}

\begin{document}
\title{Convergence of the Linear $\delta $ Expansion \\
in the Critical $O(N)$ Field Theory}
\author{Eric Braaten and Eugeniu Radescu}
\affiliation{Physics Department, Ohio State University, Columbus OH 43210, USA}

\date{\today}

\begin{abstract}
The linear $\delta$ expansion is applied to the 
3-dimensional $O(N)$ scalar field theory at its critical point
in a way that is compatible
with the large-$N$ limit. 
For a range of the arbitrary mass parameter, the linear $\delta$ expansion
for $\langle \vec \phi^{\, 2} \rangle$ converges, 
with errors decreasing like a power of the order $n$ in $\delta$. 
If the principal of minimal sensitivity is used to optimize the
convergence rate, the errors seem to decrease exponentially
with $n$.
\end{abstract}

\smallskip
\maketitle

Very few systematically improvable methods are available for
calculating nonperturbative quantities in field theory.
Such a method has the advantages that more accurate results
can be obtained at the expense of additional effort and that
reliable error estimates can be made.  One systematically improvable
nonperturbative method that is well-developed is the Monte Carlo method 
applied to the field theory formulated on a discrete lattice.
Since this method is not universally applicable, it is important to
develop other nonperturbative methods.

A classic nonperturbative problem that was recently 
solved definitively is the shift due to interactions
in the critical temperature $T_{c}$ 
for Bose-Einstein condensation (BEC).
If the potential between two bosons is short-ranged,
the leading order shift is linear in the $s$-wave scattering 
length $a$: $\Delta T_{c}/T_{c}=c \, n^{1/3}a$,
where $n$ is the number density of the bosons
and $c$ is a numerical constant. 
Baym et al.~\cite{Baym1} showed that the coefficient $c$ could be 
determined by a nonperturbative 
calculation at the critical point of an effective 
$3$-dimensional statistical field theory with $O(2)$ symmetry. 
Lattice Monte Carlo calculations
by Kashurnikov, Prokof'ev, and Svistunov
and by Arnold and Moore \cite{K-P-S}
give the result $c=1.32\pm 0.02$.
The second order correction to $\Delta T_{c}/T_{c}$ 
proportional to $( an^{1/3})^{2}$ has also been calculated
\cite{Arnold3}. 
The definitive solution to this problem makes it useful 
as a testing ground for other nonperturbative methods.

One systematically improvable nonperturbative method that has been
applied to the shift in $T_c$ is the $1/N$ expansion.
The $O(2)$ field theory relevant to BEC can be generalized to $O(N)$.
Baym, Blaizot and Zinn-Justin calculated
the coefficient $c$ analytically in the large-$N$ limit
and obtained $c=2.33$ \cite{Baym2}. 
The first correction in the $1/N$ expansion reduces $c$ to 1.72
\cite{Arnold1}.  These results seem to be converging to the 
lattice Monte Carlo result.
The analytic result in the large-$N$ limit can be used to test
other nonperturbative methods.
If a method fails to give the correct answer in the large-$N$ limit, 
one should be suspicious of its predictions for $N=2$.

Another nonperturbative method that has been applied to this problem 
is the {\it linear $\delta$ expansion} \cite{DeltaExp},
also known as {\it optimized perturbation theory} 
\cite{Stevenson:1981vj} or {\it variational perturbation theory} 
\cite{Kleinert}. 
In this method, an arbitrary parameter $m$ is introduced 
into the theory and calculations are carried
out using perturbation theory in a formal expansion 
parameter $\delta=1$. 
The rate of convergence can be improved by adjusting $m$ 
at each order in $\delta$ 
using the principal of minimal sensitivity (PMS).
It was first applied to the calculation of $\Delta T_c$
by de Souza Cruz, Pinto and Ramos \cite{Ramos}.
At $2^{\rm nd}$, $3^{\rm rd}$, and $4^{\rm th}$ order in $\delta$, 
they obtained $c=3.06$, $2.45$, and $1.48$, respectively \cite{Ramos},
which seem to be converging 
to the lattice Monte Carlo result.

The fact that corrections can be calculated systematically
is no guarantee that they actually improve the result.
Hopes for the convergence of the LDE 
are based largely on studies of its application 
to the anharmonic oscillator.
The LDE has been proven to converge for order-dependent
choices of $m$ that include the PMS criterion as a special case.
The finite temperature partition function
converges exponentially, with the errors at $n^{\rm th}$ order
decreasing as $\exp(-b Tn^{2/3}/g^{1/3})$ 
where $g$ is the strength of the anharmonic term in the potential 
and $b$ is a numerical constant \cite{Duncan-Jones}.
The energy eigenvalues $E_n$ have been proven to
converge uniformly in $g$ as $n \to \infty$ \cite{G-K-S}.
In particular, the leading term in the 
strong-coupling expansion for the ground state energy $E_0$
converges exponentially, with the errors decreasing like 
$\exp(-b' n^{1/3})$ \cite{Janke-Kleinert}.

The anharmonic oscillator is equivalent to a Euclidean field theory 
with a single real-valued field in 1 space dimension.
The statistical field theory relevant to BEC 
is a generalization to a multicomponent field in 3 space dimensions. 
The most serious obstacles to generalizing the convergence proofs 
for the anharmonic oscillator to this more complicated problem
come from the infrared (IR) and ultraviolet (UV) regions of momentum space.
It is reasonable to expect the convergence behavior to be similar if
appropriate IR and UV cutoffs are imposed 
on the field theory.  The coefficient $c$ is insensitive to 
the UV region, so we do not expect any complications
from taking the UV cutoff to $\infty$.
However, $c$ is very sensitive to the IR region, 
so convergence in the presence of an IR cutoff gives no 
information about the behavior of the LDE in the limit as the IR
cutoff goes to 0.

In this letter, we present evidence that the LDE is indeed 
a systematically improvable method.
We show that it can be implemented 
in a way that is compatible with the large-$N$ limit.
We present evidence that the coefficient $c$ in the large-$N$ limit
converges to the analytic result of Ref.~\cite{Baym2}
for a range of $m$, with errors that decrease 
as a power of the order $n$ in the LDE.
If the PMS criterion is used to optimize
the convergence rate, the errors seem to decrease exponentially in $n$.

The lagrangian density for the $O(N)$ field theory 
relevant to BEC is
\begin{equation}
{\cal L}=
-\mbox{${1 \over 2}$} \vec \phi \cdot \nabla^2 \vec \phi
+ \mbox{${1 \over 2}$} r {\vec \phi}^{\, 2}
+ \mbox{${1 \over 24}$} u \left( {\vec \phi}^{\, 2} \right)^{2},
\label{L-eff}
\end{equation}
where $\vec \phi=(\phi _{1},...,\phi _{N})$
is an $N$-component real field.
The statistical average of the operator $\vec \phi^{\, 2}$ is
\begin{equation}
\langle \vec \phi^{\, 2} \rangle =
2\int_{p}\left[p^{2}+r+\Sigma(p)\right] ^{-1},  
\label{fi2}
\end{equation}
where $\Sigma (p)$ is the self-energy 
and $\int_p = \int d^3p/(2\pi)^3$.
The integral over $\bf p$ is divergent 
and requires a UV cutoff.
The critical point can be reached by tuning the parameter $r$
to the value $r=-\Sigma (0)$.
This condition reduces to $r=0$ if $u=0$. 
The difference $\Delta$ between the critical values of
$\langle \vec \phi^{\, 2} \rangle$ at a nonzero value of $u$
and at $u=0$ is
\begin{equation}
\Delta = N \int_{p}\left[ \left[ p^{2}+\Sigma(p)-\Sigma(0)\right]
^{-1}-\left( p^{2}\right) ^{-1}\right] .  
\label{Delta-gen}
\end{equation}
The integral over $\bf p$ is convergent 
and therefore no longer requires a UV cutoff. 
At the critical point, the only relevant length scale 
is set by the parameter $u$.  Since $\Delta $ has dimensions
of length, it must be proportional to $u$ by dimensional analysis.
The determination of the coefficient of $u$ requires a 
nonperturbative calculation.  In the large-$N$ limit
defined by $N\to \infty$, $u \to 0$ with $N u$ fixed,
the coefficient is known analytically \cite{Baym2}: 
\begin{equation}
\Delta = - Nu/(96\pi ^{2})
\qquad \left( {\rm large} \, N \right).
\label{largeN-exact}
\end{equation}
The coefficient in the expression for the shift in $T_{c}$ is 
$c = -128 \pi^3 \zeta({3\over2})^{-4/3} \Delta/u$,
with $\Delta$ evaluated at $N=2$.

The {\it linear $\delta$ expansion} (LDE) is 
generated by a lagrangian whose coefficients are linear
in a formal expansion parameter $\delta$:
${\cal L}_{\delta }= (1-\delta ){\cal L}_{0} + \delta \, {\cal L}$,
where ${\cal L}_{0}$ is the lagrangian for an exactly solvable theory.
The lagrangian ${\cal L}_{\delta }$ interpolates 
between ${\cal L}_{0}$ when $\delta =0$ and ${\cal L}$ when $\delta =1$.  
To apply the LDE to the $O(N)$ statistical field theory
defined by the lagrangian (\ref{L-eff}), we choose  
the solvable field theory to be the free field theory 
with mass $m$.  The lagrangian can be written
${\cal L}_\delta = {\cal L}_{0} + {\cal L}_{\rm int}$, where
\begin{eqnarray}
{\cal L}_{0} &=&
- \mbox{${1 \over 2}$} \vec \phi \cdot \nabla^2  \vec \phi
+ \mbox{${1 \over 2}$} m^{2} \vec \phi^{\, 2},
\\
{\cal L}_{\rm int} &=&
\mbox{${1 \over 2}$} \delta  \left( r-m^{2}\right)  \vec \phi^{\, 2}
+ \mbox{${1 \over 24}$} \delta \, u \left( \vec \phi^{\, 2} \right)^{2}.
\label{L-int}
\end{eqnarray}
Calculations are carried out by expanding in powers of $\delta$,
truncating at $n^{\rm th}$ order, and then setting $\delta =1$. 

In order to apply the LDE to the shift in $T_c$,
we need a prescription for generalizing the quantity $\Delta$ 
defined in (\ref{Delta-gen}) to the field theory defined by 
the lagrangian ${\cal L}_\delta$.  
The prescription must have a 
well-defined expansion in powers of $\delta$, and it must reduce to 
(\ref{Delta-gen}) when $\delta = 1$.  The simplest prescription
is to use the expression (\ref{Delta-gen}), where $\Sigma(p)$ 
is the self-energy for the field theory with lagrangian ${\cal L}_\delta$. 
In the previous application of the LDE to the shift in $T_c$, 
the authors used an alternative prescription with the additional term 
$m^{2}(1-\delta )$ in the denominator of the first term in 
the integrand of (\ref{Delta-gen}) \cite{Ramos}. 
This prescription has the correct limit as $\delta \to 1$. 
It can be expressed as an integral with a
well-defined large-$N$ limit plus an additional term  
$-(N/4\pi)m \sqrt{1-\delta }$.
Because of this additional term, the limit $\delta\to 1$
does not commute with the large-$N$ limit. 
Using this prescription,
the prediction for $\Delta$ defined by the PMS criterion
at $n^{\rm th}$ order in the LDE, 
scales like $N^{2-1/n}u$ as $N \to \infty$, 
while the correct result is $Nu$. 

The prescription (\ref{Delta-gen})
defines $\Delta(u,m,\delta)$ as a function of 
three variables.  The $n^{\rm th}$-order approximation in the LDE is 
obtained by truncating the expansion in $\delta$ at $n^{\rm th}$ order
to obtain a function $\Delta^{(n)}(u,m,\delta)$
and then setting $\delta = 1$.
At any finite order $n$, the prediction of the LDE depends on $m$.  
As $m$ varies over its physical range from 0 to $+\infty$, 
the range of the prediction for 
$\Delta$ extends out to $\pm \infty$ depending on the order in $\delta$.  
Some prescription for $m$ is required to obtain a definite prediction.
The PMS prescription is 
$(d/dm) \Delta^{(n)}(u,m,\delta =1) =0$.
After setting $\delta =1$,
$\Delta^{(n)}$ is a function of $u$ and $m$ only.
By dimensional analysis, the value of $\Delta^{(n)}$
at a solution $m$ to the PMS criterion is proportional to $u$. 
By allowing the variable $m$ to change with the order $n$, 
the PMS criterion may improve the convergence 
rate of the LDE.

\begin{figure}[tbp]
\begin{center}
\centerline{\includegraphics[width=4cm,angle=0,clip=true]{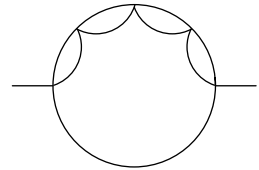}}
\end{center}
\vspace*{-18pt}
\caption{The 4$^{\rm th}$ in the series of diagrams for $\Sigma(p)$
	that survive in the large-$N$ limit.}
\label{fig:largeN}
\end{figure}

Using the prescription (\ref{Delta-gen}), the LDE for $\Delta$ 
in the large-$N$ limit can be calculated to all orders in $\delta$.
The leading contribution to 
$\Sigma(p)-\Sigma(0)$ comes from the series of diagrams 
whose $4^{\rm th}$ member is shown in Fig.~\ref{fig:largeN}.
Since $\Sigma(p)-\Sigma(0)$
is of order $1/N$, the leading contribution at large $N$
is obtained by expanding (\ref{Delta-gen})
to first order in $\Sigma(p)-\Sigma(0)$.
The expression for the large-$N$ diagram for $\Delta$ with $n+1$ loops
can be reduced to a 1-dimensional integral multiplied by $m^{-(n-1)}u^n$. 
In addition to the diagrams for $\Sigma(p)$ generated by the interaction term 
$\delta \, u (\vec \phi^{\, 2})^{2}$ in (\ref{L-int}), 
we must also take into account 
insertions of $\delta \, r$ and $-\delta \, m^{2}$. 
The effect of the $\delta \, r$ insertions is to
replace each $\Sigma(p)$ by $\Sigma(p)-\Sigma(0)$.
The effect of the $- \delta \, m^{2}$ insertions 
is to replace $m^2$ by $m^2(1-\delta)$.
Summing all the large-$N$ diagrams, we obtain
\begin{eqnarray}
\Delta &=&
\delta N u/(24 \pi^{3})
\sum_{n=2}^{\infty }
\left( - \delta / (\sqrt{1-\delta} \, \mu) \right) ^{n-1}
\nonumber
\\
&& \times \int_{0}^{\infty }dy \, y^{2} [A(y)]^{n-1}/(y^2 + 1)^{2},  
\label{D}
\end{eqnarray}
where $\mu = 48\pi m/(N u)$ and $A(y)=(2/y) \arctan(y/2)$.
The prediction for $\Delta$ at $n^{\rm th}$ order in the LDE
is obtained by expanding (\ref{D}) as a power series 
in $\delta$, truncating after order $\delta^{n}$, 
and then setting $\delta =1$. 

It is easy to show that if the LDE for (\ref{D}) converges, it
converges to the correct analytic result (\ref{largeN-exact}).  After
interchanging the order of the sum and the integral, the sum can be
evaluated. Upon taking the limit $\delta \rightarrow 1,$ 
all dependence on $\mu $ disappears.
Evaluating the integral over $y$ gives the result (\ref{largeN-exact}).
Thus, if the LDE converges at some value of $\mu$, 
it should converge to (\ref{largeN-exact}).

\begin{figure}[tbp]
\begin{center}
\centerline{\includegraphics[width=8cm,angle=0,clip=true]{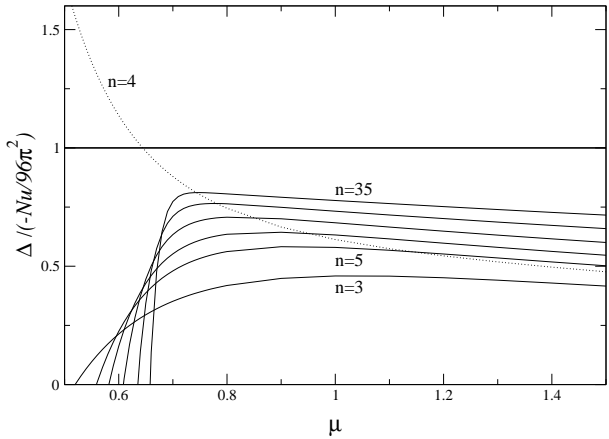}}
\end{center}
\vspace*{-18pt}
\caption{$\Delta/(-Nu/96 \pi^2)$ in the large-$N$ limit as a function of $\mu$ 
	at $n^{\rm th}$ order in the LDE for $n=3$, 4, 5, 7, 11, 19, 35.
	The curves for $n=7$, 11, and 19 
	appear in order between those labelled $n=5$ and 35.}
\label{fig:D-muN}
\end{figure}

\begin{figure}[tbp]
\begin{center}
\centerline{\includegraphics[width=8cm,angle=0,clip=true]{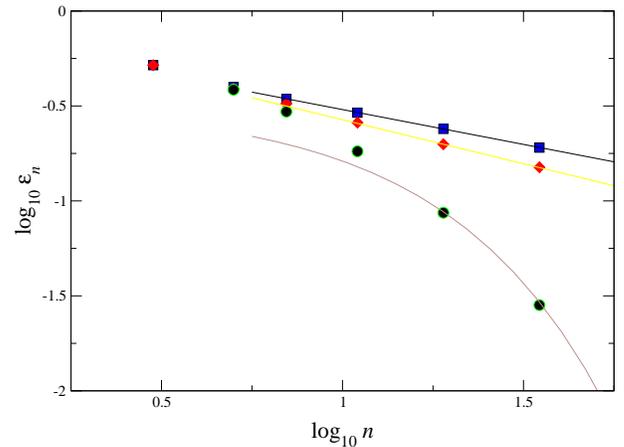}}
\end{center}
\vspace*{-18pt}
\caption{Log-log plot of the fractional error $\varepsilon _n$
	as a function of the order $n$ in the LDE.
	The squares, diamonds, and circles are for $\mu = 1.039$,
	the real solution to the PMS criterion, and the
	solution with maximal $|{\rm Im}\Delta|$, respectively.
	The lines are simple curves that pass through the last 2 points.}
\label{fig:err-n}
\end{figure}

The manipulations that showed the convergence to (\ref{largeN-exact}) 
involved several interchanges of limits. 
It is difficult to translate the conditions for the validity
of each step into a condition for the convergence 
of the LDE. However, the convergence can be easily studied numerically. 
In Fig.~\ref{fig:D-muN}, we show $\Delta $ as a function of 
$\mu $ for several orders 
in the LDE: $n=3$ and $n=2^{j}+3,$ $j=0,...,5$.
The horizontal line is the analytic result (\ref{largeN-exact}). 
The results are consistent with convergence
to (\ref{largeN-exact}) for all $\mu$ greater than a critical value 
$\mu_c$ which we estimate to be $\mu_c \approx 0.71$.
If $\mu < \mu_c$, $\Delta $
seems to diverge to $+\infty $ for $n$ even and to $-\infty $ for $n$ odd. 
For any fixed $\mu > \mu_c$, 
the convergence with $n$ is very slow. 
In Fig.~\ref{fig:err-n}, we show a log-log plot of
the fractional error $\varepsilon_n$ as a function of $n$.  
The squares lie close to a straight line, 
indicating that the errors decrease like a power of $n$.
The dotted line that goes through the last two points is 
$\varepsilon_n = 0.70n^{-0.37}$.
The errors decrease roughly like $n^{-1/3}$.

The rate of convergence can be improved by using the PMS
criterion to choose a value of $\mu$ that depends on
the order in the LDE.  
At $n^{\rm th}$ order, 
this criterion is a polynomial equation in $\mu $ of order $n-2$.
For $n$ even, there are no real roots. 
For $n$ odd, there is always one real root that corresponds 
to the maxima of the curves in Fig.~\ref{fig:D-muN}.  
The resulting fractional errors are shown as a function
of $n$ in Fig.~\ref{fig:D-muN}. 
The diamonds lie close to a straight line, 
indicating that the errors decrease like a power of $n$.
The dashed line that goes through the last two points
is $\varepsilon_n = 0.78n^{-0.46}$.
The errors decrease roughly as $n^{-1/2}$. 

\begin{figure}[tbp]
\begin{center}
\centerline{\includegraphics[width=8cm,angle=0,clip=true]{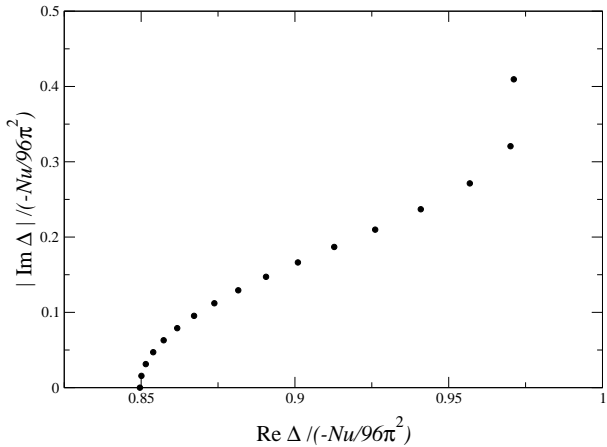}}
\end{center}
\vspace*{-18pt}
\caption{Scatter plot of $|{\rm Im}\Delta|$ vs.~Re$\Delta$ for the solutions 
	$\mu$ of the PMS criterion at $35^{\rm th}$ order in the LDE.}
\label{fig:Im-Re}
\end{figure}

Although the PMS criterion at $n^{\rm th}$ order has at most 
one real solution, there are always $n-2$ complex-valued solutions. 
Studies of the anharmonic oscillator have revealed 
that there are families of complex solutions 
with much better convergence properties than families
of real solutions \cite{B-G-N}. 
In our problem, at any odd order $n$, the real solution 
always gives the value of Re$\Delta$ that is farthest 
from the correct result (\ref{largeN-exact}). 
Thus this family of
solutions gives the slowest possible convergence rate. 
However there is a strong anticorrelation between 
the errors in Re$\Delta$ and ${\rm Im} \Delta$.
This is illustrated in Fig.~\ref{fig:Im-Re},
which is a scatter plot of $|{\rm Im}\Delta|$ vs.~Re$\Delta$ 
for the solutions to the PMS criterion for $n=35$. 
The solutions with
the most accurate values for Re$\Delta$ are those with the 
largest values for $| {\rm Im} \Delta|$.
Thus we can define a nearly optimal family of solutions by choosing 
those with the maximal values of $| {\rm Im} \Delta|$.
While Im$\Delta$ for these solutions shows no sign of converging 
to $0$, there is dramatic improvement in the convergence 
of Re$\Delta$. 
A log-log plot of the fractional errors in Re$\Delta$ 
is shown in Fig.~\ref{fig:D-muN}. 
The downward curvature of the points
indicates that the errors decrease faster than any power of $n$.
The solid line that goes through the last two points is 
$\varepsilon_n = 0.32 (0.93)^n$.
The errors are decreasing faster than this exponential.

We have calculated $\Delta$ for finite $N$ 
to $4^{\rm th}$ order in the LDE 
using our prescription \cite{Braaten-Radescu}.
Setting $N=2$ and using the PMS criterion, 
we obtain a real result $\Delta^{(3)} = 0.192$ at 
$3^{\rm rd}$ order and a complex result 
$\Delta^{(4)} = 0.214\pm 0.084 i$ at $4^{\rm th}$ order.
Taking $\Delta^{(3)}$ and Re$\Delta^{(4)}$ as the predictions,
the corresponding values for the coefficient in $\Delta T_c$
are $c=0.447$ and 0.492.
They seem to be slowly approaching the lattice Monte Carlo result 
$ c=1.32 \pm 0.02$ \cite{K-P-S} from below.
The errors in the $3^{\rm rd}$ and $4^{\rm th}$ order results are
66\% and 62\%, which is larger than the errors of 
52\% and 44\% in the corresponding predictions in the large-$N$ limit.
The percentage improvement in going from $3^{\rm rd}$ 
to $4^{\rm th}$ order is only half as large as in the 
large-$N$ limit.  The $3^{\rm rd}$ and $4^{\rm th}$ order
predictions for $N=2$ using the prescription of Ref.~\cite{Ramos}
have errors of 85\% and 20\%.  
The small error in their $4^{\rm th}$ order prediction 
may be fortuitous.  The low order predictions using their
prescription are skewed by the term in 
$\Delta$ that does not have a well-behaved large-$N$ limit.
The prescription for $\Delta$ obtained by deleting that term 
is equally valid, and it gives values below the lattice Monte Carlo result
with errors of 68\% and 65\%.

Although the convergence rate of the optimized LDE appears to be 
exponential in the large-$N$ limit, it is still rather slow. 
One must calculate to about $18^{\rm th}$ order in $\delta$ 
to achieve 10\% accuracy.  For general $N$, it may be feasible to 
calculate $\Delta$ to $5^{\rm th}$ order,
but it would be difficult to go to much higher order.  
The slow convergence in the large-$N$ limit
suggests that even if the LDE also converges for $N=2$, 
a strict expansion in $\delta$ is not useful for quantitative 
calculations.  It may however be possible to use order-dependent 
mappings to change the expansion in $\delta$ into 
a more rapidly converging expansion \cite{Seznec:ev}.

In conclusion, we have shown that the LDE for the quantity $\Delta$ 
in the 3-dimensional critical $O(N)$ field theory
in the large $N$ limit converges if we use an appropriate prescription.
If the PMS criterion is used to optimize the convergence,
the errors seem to decrease exponentially in the order of the LDE.
This provides some hope that the LDE may also be a systematically improvable 
approximation scheme in the case $N=2$ relevant to BEC.
With the use of order-dependent mappings to accelerate the convergence, 
it may be possible to develop the LDE into a general and powerful tool for 
quantitative calculations in superrenormalizable field
theories, even at a critical point.

This research was supported in part by DOE grant DE-FG02-91-ER4069.

\end{document}